# Kinematic and Energetic Properties of the 2012 March 12 Polar Coronal Mass Ejection

N. Gopalswamy

Solar Physics Laboratory, NASA Goddard Space Flight Center, Greenbelt, MD 20771

nat.gopalswamy@nasa.gov

S. Yashiro[1] and S. Akiyama[1]

The Catholic University of America, Washington, DC 20064

[1] also at NASA Goddard Space Flight Center, Greenbelt, MD 20771

## ABSTRACT

We report on the energetics of the 2012 March 12 polar coronal mass ejection (CME) originating from a southern latitude of ~60º. The polar CME is similar to low-latitude CMEs in almost all respects: three-part morphology, post eruption arcade (PEA), CME and filament kinematics, CME mass and kinetic energy, and the relative thermal energy content of the PEA. From polarized brightness images, we estimate the CME mass, which is close to the average mass of low-latitude CMEs. The CME kinetic energy ($3.3 \times 10^{30}$ erg) is also typical of the general population of CMEs. From photospheric magnetograms, we estimate the free energy ($1.8 \times 10^{31}$ erg) in the polar crown source region, which we find is sufficient to power the CME and the PEA. About 19% of the free energy went into the CME kinetic energy. We compute the thermal energy content of the PEA ($2.3 \times 10^{29}$ erg) and find it to be a small fraction (6.8%) of the CME kinetic energy. This fraction is remarkably similar to that in active region CMEs associated with major flares. We also show that the 2012 March 12 is one among scores of polar CMEs observed during the maximum phase of cycle 24. The cycle 24 polar crown prominence eruptions have the same rate of association with CMEs as those from low-latitudes. This investigation supports the view that all CMEs are magnetically propelled from closed field regions, irrespective of their location on the Sun (polar crown filament regions, quiescent filament regions or active regions).

**Subject headings**: Sun: coronal mass ejections - Sun: filaments - Sun: prominences - Sun: flares - Sun: magnetic fields

## 1. Introduction

There are two types of mass emission from the Sun: the ubiquitous solar wind and coronal mass ejections (CMEs), which are transient. There are also small-scale structures that seem to accelerate similar to the slow solar wind, unlike CMEs. Uchida et al. (1992) identified active-region expansion in Yohkoh soft X-ray images that were slow, typically of the order of a few to a few tens of km/s. Uchida et al. (1992) did not find these expansions to be associated with any reconnection. Gosling (1990) discussed interplanetary extension of rising loops from the corona with no reconnection occurring near the Sun. Such structures were proposed to explain interplanetary CMEs (ICMEs) that do not have a flux rope structure (i.e., magnetic clouds). For example, such loops would indicate counter-streaming superthermal electron flux, one of the signatures of an interplanetary CME (ICME). On the other hand when a CME is produced due to reconnection, one would observe a flux rope ICME in the interplanetary medium and post eruption arcades (PEAs) at the Sun. Recent investigations of heavy element charge states inside ICMEs have shown that both magnetic-cloud and non-cloud interplanetary CMEs (ICMEs) show enhanced charge states suggesting that both types of ICMEs might have originated from flare reconnection (Gopalswamy et al. 2013). This was further confirmed by the fact that the solar sources of both types of ICMEs have PEAs (Yashiro et al. 2013). Sheeley et al. (1997) identified blobs originating in helmet streamers at a heliocentric distance of about 3-4 Rs, accelerate slowly (~4 m/s$^2$) from ~150 km/s at 5 Rs to 300 km/s at ~25 Rs. They concluded that the blob acceleration is similar to the slow solar wind expansion. Antiochos et al. (1999) suggested that CMEs associated with polar crown filaments (PCFs) should not be considered as CMEs, but merely coronal mass expansions. They further suggested that at a few solar radii, the plasma dominates the field and expands outward indefinitely, similar to the Sheeley et al. (1997) events.

Recently, Gopalswamy (2013; 2015) reported on the polar CME of 2012 March 12 associated with a PCF located at a latitude of ~S60. The CME has all the classical features like any other CME associated with a prominence eruption (PE) demonstrating the following points. 1. The polar CME also has the three-part morphology (Hundhausen 1987). 2. The polar CME originates in a helmet streamer overlying a PCF. 3. A PEA is formed during the polar CME eruption, with its feet located on either side of the pre-eruption location of the filament (two-ribbon structure). 4. The CME speed in the field of view (FOV) of SOHO's Large Angle and Spectrometric Coronagraph (LASCO) is slightly above the average value (~475 km/s) of the general population of CMEs. 5. The peak value of the initial acceleration of the polar CME is typical of prominence-associated CMEs (>100 m/s$^2$). 6. The polar CME attains its peak acceleration at a height of ~2.5 Rs, which is typical of most low-latitude CMEs (see, e.g., Bein et al. 2011).

The common feature of normal CMEs is the three-part structure (plus shock for fast CMEs) that is ejected from the Sun, leaving behind a PEA straddling the source polarity inversion line. The PEA is one of the prominent near-surface signatures of eruptions indicating a close relation between flares and CMEs. There are several other properties that confirm the close connection between flares and CMEs: synchronous flare-CME acceleration (Zhang et al. 2001; Zhang and Dere, 2006), magnetic flux equality between flare reconnection flux and magnetic cloud azimuthal flux (Qiu et al. 2007), flare magnetic field – CME width relationship (Moore et al. 2007), spatial relationship between CME direction and flare location (Yashiro et al. 2008), soft X-ray flare fluence – CME kinetic energy relationship (Gopalswamy 2010), flare temperature – ICME charge state relationship (Gopalswamy et al. 2013; Yashiro et al. 2013), and so on. If the polar crown CMEs follow these established relationships, one can gain a unified understanding of the eruption process as a common mechanism on the Sun, irrespective of the source latitude.

The purpose of this study is to quantitatively determine the energetics of the 2012 March 12 polar CME and the associated PEA. We also determine the magnetic potential energy in the source region (as a proxy to the free energy available) and compare it with the CME kinetic energy, and the thermal energy content of the PEA. We also show that the 2012 March 12 polar CME is typical of a large number of polar CMEs observed during the maximum phase of solar cycle 24 by deriving the speed, width, and acceleration of CMEs associated with PEs at high and low latitudes.

## 2. Observations

The 2012 March 12 CME associated with a polar crown filament eruption was observed by the coronagraphs on board the Solar and Heliospheric Observatory (SOHO) and the Solar Terrestrial Relations Observatory (STEREO) missions (Brueckner et al. 1995; Howard et al. 2008). The filament eruption and the PEAs were observed by the Extreme Ultra Violet Imager (EUVI, Howard et al. 2008) on board STEREO and the Atmospheric Imaging Assembly (AIA, Lemen et al. 2012) on board the Solar Dynamics Observatory (SDO). The prominence eruption (PE) was also imaged in microwaves using the Nobeyama Radioheliograph (Nakajima et al., 1994). The PE in SDO/AIA 340 Å and the microwave images were consistent (Gopalswamy, 2015). Gopalswamy (2015) reported on the initial evolution of the CME and the associated PE. In particular, he showed that the CME was observed with its leading edge ahead of the eruptive prominence with a void in between, similar to the familiar three-part structure observed in regular CMEs: the leading edge, the dark void, and the prominence core,(Hundhausen, 1987). Here we focus on the CME observations in the outer corona and compare them with the near-surface results reported in Gopalswamy (2015).

Figure 1 summarizes the observations reported in Gopalswamy (2015) to provide context to the investigation presented in this paper. The height-time measurements of the 2012 March 12 prominence, CME, and the prominence core using SDO and STEREO data show that the EUV prominence becomes the prominence core observed by STEREO/COR1. Within the COR1 FOV (1.4 – 4 Rs), the CME leading edge attains a speed of ~400 km/s (Fig. 1e). The initial peak acceleration of the CME in the COR1 FOV was more than 150 m/s$^2$ for both the leading edge and the prominence core (Fig. 1f). Fitting a second order polynomial to the CME height-time plot, we get an average acceleration of 51.5 m/s$^2$ within the COR1 FOV. This acceleration is more than an order of magnitude larger than the acceleration of the Sheeley blobs (Sheeley et al.

1997). Figure 1 also shows that the PEA intensity ($I$) in EUV resembles a gradual soft X-ray event except that the intensity is very low. The increase in $I$ closely follows the increase in CME speed, illustrating the close connection between CMEs and flares (Zhang et al., 2001). The time derivative of the intensity ($dI/dt$) peaks in the rise phase of $I$, mimicking the Neupert effect (Neupert, 1968; Dennis & Zarro, 1993). The $dI/dt$ peak also roughly coincides with the peak acceleration of the CME leading edge and the prominence core. The peak acceleration occurs when the CME reaches a height of ~2.5 Rs as in low-latitude CMEs (Wood et al., 1999; Bein et al. 2011).

## 3. CME Kinematics in the Outer Corona

Figure 2 compares the CME images from STEREO/COR1 and COR2 and the continued acceleration of the CME in the COR2 FOV. The eruption was front-sided in STEREO-Behind (STB) view, as can be seen from the lifting filament becoming eruptive prominence in Fig. 2(b). The filament maintained its shape as it became the CME core in the STB/COR1 image, which also shows the void and frontal structure of the CME. In Fig. 2(c), the CME has expanded to a larger size, but all the substructures can be clearly recognized. Figure 2(d) shows the height-time plot of the CME as it traversed the STB/COR2 FOV. We have also combined the height-time measurements from EUVI, COR1 (shown in Fig. 1) and COR2 to get the full evolution of the CME. The height-time plots of the leading edge (upper curve) and the prominence core (lower curve) are compared. In this view, the time resolution is ~1 h, so we do not see the finer details given in Fig.1. The height-time plot clearly shows that the filament accelerated slowly in the beginning (7.1 ms$^{-2}$), followed by a rapid acceleration in the COR1 FOV (45.7 ms$^{-2}$), and again lower acceleration in the COR2 FOV (9.6 ms$^{-2}$). The leading edge had a similar behavior, except that the magnitude of acceleration was slightly higher: 51.5 ms$^{-2}$ in the COR1 FOV and 22.3 ms$^{-2}$ in the COR2 FOV. Thus in the outer coronagraph FOV, the CME continued to accelerate, a behavior typical of most CMEs associated with PEs (see Gopalswamy et al. 2003a).

Figure 3 shows the height-time history of the CME as viewed from the Sun-Earth line (prominence from SDO/AIA at 304 Å; CME and the prominence core from SOHO/LASCO C2 and C3). These observations are generally consistent with the STB observations with the three stages of acceleration: slow, fast, and slow. The acceleration values are slightly different because of the slightly different projection effects and the larger LASCO FOV (32 Rs vs. 15 Rs in STB/COR2). The maximum acceleration of the CME (86.7 ms$^{-2}$) and the prominence (20.1 ms$^{-2}$) were in the LASCO/C2 FOV. The last measurement was made in the LASCO/C3 FOV at 08:18 UT when the CME was at a heliocentric distance of ~23.75 Rs. The CME reached a speed of ~750 km/s near the outer edge of the LASCO/C3 FOV, while the average speed within the combined C2 and C3 FOV was ~640 km/s. The average CME speed is near the high end of the range of speeds obtained for high-latitude CMEs in cycle 23 (Gopalswamy 2015) measured in the LASCO FOV (no measurements in the inner corona for cycle-23 CMEs because LASCO/C1 stopped functioning before the appearance of polar-crown CMEs). The width of the 2012 March 12 CME (122°) in the LASCO FOV was also near the high end of the width range (http://cdaw.gsfc.nasa.gov/CME_list/UNIVERSAL/2012_03/univ2012_03.html).

## 4. CME Mass and Kinetic Energy

The SOHO/LASCO CME catalog lists mass and kinetic energy of CMEs that have widths in the range 20 to 120°. The CME in question was a partial halo (width ~122°) and hence was not included in the catalog. Therefore, we estimated the mass from the lone polarized brightness (*pB*) image obtained when the CME was in the C2 FOV. Figure 4 shows *pB* images of the pre-event corona and the CME. A difference image is also shown with the excess brightness due to the CME outlined by a single contour. Note that the pre-event image was obtained about 6 hours before the event image.

We computed the CME mass by estimating the number of electrons within the outlined area in Fig. 4c required to produce the observed *pB* and then multiplying it by the mass associated with each electron ($1.97 \times 10^{-24}$ g/e) assuming 10% helium. We estimated $8.15 \times 10^{38}$ electrons, resulting in a CME mass of $1.6 \times 10^{15}$ g when all the electrons are assumed to be in the plane of the sky. If the electrons are not in the sky plane, the mass estimated increases with increasing angle from the sky plane. For example, if the electrons are located in a plane that is 30° away from the sky plane, the CME mass becomes $2.8 \times 10^{15}$ g, which is only 75% higher than the sky-plane case. It must be pointed out that the above mass is likely to be an underestimate because for regular CMEs, the mass increases up to about 7 Rs and then attains a quasi-steady value. However, the CME was at a height of ~6 Rs at 02:58 UT, when the *pB* image was obtained. Therefore, the final mass may not be too much higher than the one computed above. The computed mass is typical of regular CMEs ($\sim 1.3 \times 10^{15}$ g) that originate near the limb (see e.g. Gopalswamy et al. 2010). The empirical relation between CME mass (*M*) and width (*W*) derived by Gopalswamy et al. (2005a), viz., *log M = 12.6 + 1.3logW* also gives a mass of $\sim 2 \times 10^{15}$ g suggesting that there is nothing unusual about the 2012 March 12 CME. Gopalswamy et al. (2015a) reported that the average CME mass ($1.1 \times 10^{15}$ g) in cycle 24 is smaller than that ($3.1 \times 10^{15}$ g) in cycle 23 for a set of limb CMEs associated with flares of size C3 and larger. The mass of the 2012 March 12 CME is above average compared to the cycle-24 masses.

Using the average CME speed (640 km/s) in the LASCO FOV and the mass derived above ($1.6 \times 10^{15}$ g), we get a kinetic energy of ~ $3.3 \times 10^{30}$ erg. Using the maximum speed (746 km/s) within the LASCO FOV, the kinetic energy becomes $4.5 \times 10^{30}$ erg. These numbers are slightly larger than the average kinetic energy of limb CMEs in cycle 23 ($1.6 \times 10^{30}$ erg – see Gopalswamy et al. 2010). We now check if this energy can be supplied by the free energy in the source region.

## 5 Magnetic potential Energy in the Polar CME Source Region

In order to estimate the magnetic potential energy of the source region, we use the synoptic charts of the radial magnetic field from SDO's Helioseismic and Magnetic Imager (HMI) data.

According to the estimate by Mackay et al. (1997), the free energy in the source region is roughly equal to the magnetic potential energy. We need the average field strength ($B_s$) in the source region, and the volume of the source region to estimate the magnetic potential energy. We estimate $B_s$ from the HMI synoptic chart. We assume the area of the source region to be the same as the maximum area ($A_s$) of the PEA. The volume is then estimated as $A_s^{3/2}$. Note that this may a lead to an underestimate of the actual volume because the field lines occupy more volume in the corona. However, this estimate is sufficient for the present purposes.

Figure 5 shows the HMI synoptic chart for the 2012 March 12 event. We estimate an average $B_s$ within the quadrilateral in Fig. 5 (negative polarity side of the bipolar source region) and then assume that the same average field is present over the whole area under the arcade. We do this because the positive polarity side is too close to the pole and the magnetic field estimate is not very accurate. The average of unsigned $B_s$ is ~ 5.47 G. The area under the arcade is obtained from the longitudinal and latitudinal extents of the PEA: the longitudinal extent (E113 to E41) corresponds to a length of $2.8 \times 10^{10}$ cm and the width of the arcade is $2.2 \times 10^{10}$ cm, giving $A_s = 6.16 \times 10^{20}$ cm$^2$ and the volume $V_s = 1.52 \times 10^{31}$ cm$^3$. Alternatively, we can assume that the height of the arcade is equal to half of the width to get $V_s$ as ~ $6.6 \times 10^{30}$ cm$^3$. Since PEAs form only over part of the source region, we think this volume is an underestimate and use $1.52 \times 10^{31}$ cm$^3$. The magnetic potential energy is then $B_s^2 V_s / 8\pi = 1.8 \times 10^{31}$ erg. We see that the CME kinetic energy ($3.3 \times 10^{30}$ erg) is less than the magnetic potential energy. In other words, only ~ 19% of the available free energy is converted into CME kinetic energy. This fraction is typical of low-latitude CMEs from active regions (see e.g., Gopalswamy et al. 2005b). Thus the polar CME of 2012 March 12 has its kinematics, energetics, and flare connection very similar to those of normal CMEs from lower latitudes.

We now discuss whether the free energy estimated above is consistent with the range of initial acceleration observed in the 2012 March 12 CME. The theoretical basis for the initial acceleration is the free energy available in the source region (Vršnak 2008; Gopalswamy et al. 2010). Assuming that all the free energy built up in the source region goes into the CME kinetic energy, one can write

$$1/2 \rho V^2 \leq B_s^2 / 8\pi, \qquad (1)$$

where $\rho$ is the plasma density and $B_s$ is the average magnetic field in the source region, and $V$ is the CME speed. Since $V_A = B/(4\pi\rho)^{1/2}$ is the Alfvén speed in the source-region corona, this relation simplifies to

$$V \leq V_A. \qquad (2)$$

If the initial size of the CME is $S$, the Alfvén transit time $\tau = S/V_A$, which when combined with the CME speed $V$ gives the initial acceleration:

$$a = V/\tau \leq V_A^2 / S. \qquad (3)$$

For the observed $a = 0.15$ km/s$^2$ and $S$ taken as the typical height of the polar coronal cavity (0.6 Rs or $S = 4.2 \times 10^{10}$ cm), one gets $V_A \geq 250$ km/s, which is readily satisfied (Mann et al. 1999; Gopalswamy et al. 2001). We can show this from the estimated $B_s = 5.47$ G for the 2012 March 12 CME: a coronal density of ~$2 \times 10^9$ cm$^{-3}$ in the inner corona, we get $V_A = 266$ km/s. Accelerations up to 10 kms$^{-2}$ have been estimated for very fast CMEs (Zhang and Dere, 2006; Bein et al. 2011; Gopalswamy et al. 2012a). Such high values can be readily accounted for in active region sources where the free energy as high as $10^{33}$ erg has been found (Gopalswamy et al. 2005b).

**6 Flare Temperature and Thermal Energy Content**

In order to further confirm the PEA – CME relationship, we determine the temperature and thermal energy content of the PEA. It is known from lower-latitude eruptions that the PEA temperature is in the range 5-30 MK (Gopalswamy et al. 2013; Yashiro et al. 2013). We shall use the technique developed by Aschwanden et al. (2013) that makes use of the six SDO/AIA filter responses to obtain the peak temperature of the arcade (see also Battaglia and Kontar 2012). Aschwanden et al. (2013) showed that the temperature and emission measure can be obtained in various coronal structures from the coronal holes to active regions, so long as the temperatures are in the range 0.5 – 10 MK. The differential emission measure (DEM) distribution is modeled by a Gaussian function in logarithmic temperature. By fitting Gaussian functions to the observations, one can obtain the peak DEM and temperature in the coronal structure of interest. This technique is readily applicable to the PEAs because they are in the temperature range of applicability (between coronal holes and active region flares). We obtained the temperature maps using the following steps: 1) make 2048 x 2048 binned images (2 x 2 macro pixel) from the original images obtained in each of the 6 filters (94, 131, 171, 193, 211, and 335 Å); 2) cut out a 512 x 512 sub-region that includes the PEA; 3) obtain the DEM map assuming a Gaussian DEM distribution and applying the empirical multiplier of 4.9 for the 94 Å image (see Aschwanden et al. 2013 for details); 4) get the peak Temperature ($T_P$), width of the temperature distribution ($\sigma_T$), and the peak Emission Measure ($EM_P$) by minimizing $\chi^2$; 5) for each pixel, calculate the emission measure using $T_P$, $\sigma_T$, and $EM_P$. The temperature map obtained this way is shown in Fig. 6. It is clear that the PEA appears in the source region and is heated to ~2 MK (Fig. 6c), compared to the surrounding corona at ~1 MK (Fig. 6a). Such heating has been inferred in low-latitude PEAs from microwave and soft X-ray images (Hanaoka et al. 1994; Gopalswamy et al. 2013).

Finally, we use the temperature and emission measure analysis to estimate the thermal energy content in the PEAs to compare it with the CME kinetic energy derived from coronagraphic observations. For each pixel, we calculated the coronal density $n = (EM/L)^{\frac{1}{2}}$ and the thermal energy density $3nk_bT_P$, where $L$ is the line-of-sight depth of the arcade (assumed to be ~$10^9$ cm) and $k_b$ is the Boltzmann's constant. 2) The thermal energy of the arcade is given by

$$U_{th} = \sum_{i=0}^{N-1} 3\, n_i\, k_b\, T_{Pi}\, AL, \qquad (4)$$

where $A$ is the pixel area ($7.5 \times 10^{15}$ cm$^2$), $i$ is the pixel index, $N$ is the total number of pixels within the PEA area (enclosed by the polygon in Fig. 5(c); $N = 26467$). The total area of the polygon is $1.99 \times 10^{20}$ cm$^2$. The resulting peak thermal energy is shown in Fig. 7 as a function of time. The maximum value of the thermal energy content of the arcade is $\sim 7.3 \times 10^{28}$ erg. From STB/EUVI observations (see Fig. 5), we know that the arcade extends behind the east limb, so the total area of the arcade is (see section 5) $6.1 \times 10^{20}$ cm$^2$, which is larger than the area of the polygon by a factor of 3.1. We can correct for this by simply multiplying the thermal energy content obtained from equation (4) by a factor of 3.1, assuming that the arcade is approximately uniform over the full extent shown in Fig. 5. Thus the estimated thermal energy content is $\sim 2.26 \times 10^{29}$ erg. Recall that the associated CME had a kinetic energy of $3.3 \times 10^{30}$ erg, so the peak thermal energy is $\sim 6.8\%$ of the CME kinetic energy. On the other hand the peak thermal energy is only $\sim 1.2\%$ of the free energy available in the polar CME source region.

How does the ratio of thermal energy to CME kinetic energy for the polar CME compare with that from low-latitude eruptions? Emslie et al. (2012) estimated the peak thermal energy and CME kinetic energy in a set of eruptions associated with major flares (M and X class) that produced solar energetic particle events. From their Table 1, we found 17 events with estimated peak thermal energy and CME kinetic energy, from which we computed the ratio in question. The ratio is in the range 0.5% to 10% with mean and median values of 3.2% and 2.8%, respectively. Our value (6.8%) is in remarkable agreement with the range of values from Emslie et al. (2012), even though the thermal energy content of the PEA is smaller than the low-latitude values by two orders of magnitude. Therefore, we can safely conclude that even from the energetics point of view, the weak flare-like brightening associated with the polar crown CME is no different from the major flares from active regions.

## 7. Other Polar CMEs

The polar CME presented above is one of more than 100 PEs from the north and south polar regions during cycle 24 (up to the end of 2014). In Figure 8, we have plotted the source centroids of all PEs detected automatically using the SDO/AIA data with the false detections eliminated by careful checking. We have also plotted the tilt angle of the heliospheric current sheet obtained by the Wilcox Solar Observatory (http://wso.stanford.edu/Tilts.html). The high latitude (HL) PEs occur when the tilt angle crosses the 60º latitude. The tilt angle sharply increases to high values before the end of 2010 in the northern hemisphere, while the increase is gradual and crosses 60º only in the beginning of 2012 in the southern hemisphere. The tilt angle has started becoming smaller in 2015, so we do not expect more HL PEs. The polar PEs occur mainly during the solar maximum phase and their cessation marks the epoch of polarity reversal at solar poles (Hyder, 1965; Gopalswamy et al. 2003b; Shimojo et al. 2006; Gopalswamy et al. 2012b).

## 7.1 CME Association Rate of PEs

The most common solar phenomenon associated with CMEs are PEs (Munro et al. 1979; Webb and Hundhausen, 1987; St. Cyr and Webb, 1991). Munro et al. (1979) found that 70% of CMEs are associated with PEs. Studies starting with PEs and connecting them to CMEs have shown that ~72% of PEs had CME association (Gilbert et al. 2000; Hori and Culhane, 2002; Gopalswamy et al. 2003a). We confirm these statistical results using the polar PE data for cycle 24. Figure 8 shows that the polar PEs are very frequent during the maximum phase: there were 108 PEs from latitudes ≥60º during cycle 24 until the end of 2014. We examined the CME association of the PEs using SOHO and STEREO coronagraph data. We found that 76 of the 108 polar PEs (or 70%) were associated with white light CMEs. It is remarkable that the association rate is nearly the same as that (72%) of all PEs reported in earlier cycles as noted above. For comparison, we also examined the CME association rate of PEs from low latitudes (within ±40º) in cycle 24. Out of the 404 low latitude (LL) PEs, 286 were associated with white-light CMEs. This again corresponds to an association rate of 71%, very similar to the HL CME association rate with PEs.

## 7.2 LL and HL CME Properties

Figure 9 compares the HL and LL CME properties in cycle 24. The CME speed distributions are very similar except for a set of LL CMEs that had speed >1000 km/s. The HL CMEs had speeds up to ~800 km/s (average speed within the coronagraph FOV). In the case of LL CMEs, the speeds were as high as 1750 km/s. The average speed of the LL CMEs is about 29% higher than that of the HL CMEs. The difference is larger than the typical measurement errors (~10%). The range of widths is similar for HL and LL CMEs, although the number of wider CMEs is larger in the LL population. There are no full halo CMEs in either population. The average width of LL CMEs (110º) is higher than that of HL CMEs (58º) by ~90%. Finally, the acceleration distributions of HL and LL CMEs show an average positive acceleration, which is characteristic of CMEs associated with PEs. We performed the Kolmogorov-Smirnov test to check if the HL and LL distributions are different. The test confirmed that all the three distributions are significantly different between high and low latitudes. The probability (p) that the distributions are different by chance is very low in each case: 0.003 (speed), 0.000 (width), and 0.012 (acceleration). We think the difference between the HL and LL populations is understandable because the LL CME source regions have higher magnetic field strength and hence higher free energy available for powering the CMEs, as reflected in the speeds. Furthermore, PEs from active regions are likely to be included in the LL population by the automatic detection algorithm. Faster CMEs are known to be wider, which explains the higher width of LL CMEs. Finally, the acceleration in Fig. 9 is the residual acceleration in the coronagraph FOV. The residual acceleration is positive for most of the events because the propelling force acts over a larger heliocentric distance for PE-associated CMEs. The smaller acceleration of LL CMEs reflects the fact that faster CMEs are subject to more drag.

## 8. Discussion

We studied the source region, kinematics, and energetics of the 2012 March 12 south polar-crown CME using observations from multiple spacecraft (SOHO, STEREO, and SDO). The prominence was tracked in EUV images obtained by the two STEREO spacecraft and the SDO/AIA when it is close to the Sun. The prominence became the CME core as observed in the inner and outer coronagraphs of the two STEREO spacecraft and LASCO C2 and C3. It was possible to determine the mass of the CME using *pB* images and hence the kinetic energy. The polar CME eruption resulted in a large PEA whose thermal energy content was a small fraction of the CME kinetic energy. The magnetic potential energy was sufficient to power the CME and produce the PEA. We also showed that the 2012 March 12 event is one of the large number of polar CMEs observed by these spacecraft. The 2012 March 12 CME had an above-average speed (640 km/s) and width (122º), while the residual acceleration (7.6 ms$^{-2}$) was slightly below the average acceleration of the polar-crown CMEs. The observations presented here confirm that the polar-crown CMEs are similar to their low-latitude counterparts in many respects.

The peak initial acceleration of the 2012 March 12 CME was ~150 ms$^{-2}$, which is similar to that of low-latitude CMEs associated with filament eruptions outside active regions including those associated with large solar energetic particle (SEP) events (Kahler et al. 1986; Gopalswamy et al. 2015b). The initial accelerations of CMEs associated with filament eruptions fall at the lower end of the initial acceleration distribution of all CMEs. The main difference between these SEP events from quiescent filament regions and the ones from active regions is that the former have a very steep spectrum (spectral index ≥4) in the 10-100 MeV energy range. On the other hand, SEP events with ground level enhancement (GLE) have a harder spectrum (spectral index ~2). The difference seems to stem from the impulsive acceleration in GLE CMEs and a more gradual acceleration in FE CMEs resulting in small and large shock formation heights, respectively in the corona.

The different initial accelerations of PE-related and flare-related CMEs have been discussed several times in the past (see e.g., Moon et al. 2002; Chen and Krall, 2003 and references therein). The primary difference seems to be that the flare-related CMEs have a larger average speed in the coronagraph FOV than the PE-related CMEs. This difference is essentially due to the difference in the magnitude of initial acceleration in the two cases. Furthermore, if the weak PEAs are included, almost all CMEs can be considered flare related and there is a continuum of acceleration magnitudes starting from slow acceleration on the one end and the impulsive, large acceleration in GLE events. The range of initial accelerations can be related to the free magnetic energy available in the source region as was discussed in section 5. Studying the acceleration of CMEs, Chen and Krall (2003) concluded that one mechanism is sufficient to explain flare-related and prominence-related CMEs. This can also be seen from the fact that equation (3) covers the entire range of accelerations from filament eruption CMEs to major flare-related ones.

One of the important implications of this study is that large-scale eruptions with CME and flare signatures commonly occur in bipolar magnetic regions. Polar crown filament regions and quiescent filament regions at low latitudes are good examples of purely bipolar regions. The magnetic energy stored in such bipolar regions seem to be sufficient to power CMEs from these regions. Small-scale flux emergence with appropriate orientation near the filaments seems to trigger the eruption from such regions as shown from observations (Feynman and Martin 1995, Gopalswamy 1999, Marqué 2004; Gopalswamy et al. 2006) and simulations (Chen and Shibata 2000; Kusano et al. 2012). The observations presented here thus support theories of CME initiation in bipolar regions. Such initiation mechanism may also be relevant to complex active regions because flux emergence and filament eruption also take place in active regions. Recently, Kleint et al. (2015) presented evidence that the filament destabilization and eruption are closely related to flux emergence and cancellation occurring around the pores in an active region. These observations confirm that reconnection-favoring flux emergence is the key to the eruption, irrespective of the source structure (bipolar or multipolar) and hence the mechanism has universal appeal, consistent with the standard bipolar model for eruptive flares (Moore et al. 2001; Sterling and Moore, 2003; Moore & Sterling, 2006). Models requiring multipolar structure in the source region (Antiochos et al. 1999) may not be applicable to CMEs originating in bipolar regions, such as the ones presented in this paper.

## 9. Conclusions

The primary result of this paper is that the magnetic free energy available in the polar crown filament region is more than sufficient to power the polar CME on 2012 March 12 and the associated PEA. From EUV images, we determined the temperature of the PEA, clearly increasing above the quiet Sun level as the eruption proceeded. In addition to the morphological and kinematic similarity between polar CMEs and low latitude CMEs, we find similarity in the energy budget. In particular we find that the fraction of free energy going into mass motion (CME) and plasma heating (PEA) in polar crown eruptions is similar to that in low-latitude eruptions from active regions. The CME kinetic energy is ~19% of the estimated free energy in the source region. The thermal energy content of the PEA is only a small fraction of the CME kinetic energy (6.8%) and the free energy (1.2%). The ratio of the peak thermal energy to the CME kinetic energy is similar in polar crown eruptions and active region eruptions. We also found that the 2012 March 12 CME is one among scores of polar CMEs. We compared the statistical properties of CMEs associated with polar and low-latitude eruptions. There are some significant differences between the polar and low-latitude populations, which simply reflect the different amounts of free energy available. The range of initial accelerations in the combined set of polar and low-latitude eruptions can be readily accounted for using the free energy available in the source regions, thus supporting a unified eruption mechanism in quiescent filament regions (including polar crown filament regions) and active regions.

**Acknowledgements**

This research benefited from NASA's open data policy in making the SOHO, STEREO, and SDO data available on line. This work was supported by NASA's LWS TR&T program.**References**

Antiochos, S. K., Devore, C. R., and Klimchuk, J. A. 1999, A Model for Solar Coronal Mass Ejections, ApJ, 510, 485

Aschwanden, M. J., Boerner, P., Schrijver, C. J., Malanushenko, A. 2013, Automated Temperature and Emission Measure Analysis of Coronal Loops and Active Regions Observed with the Atmospheric Imaging Assembly on the Solar Dynamics Observatory (SDO/AIA), Solar Phys., 283, 5

Battaglia, M., Kontar, E. P. 2012, RHESSI and SDO/AIA Observations of the Chromospheric and Coronal Plasma Parameters during a Solar Flare, ApJ, 760, 142

Bein, B. M., Berkebile-Stoiser, S., Veronig, A. M., et al., 2011, Impulsive Acceleration of Coronal Mass Ejections. I. Statistics and Coronal Mass Ejection Source Region Characteristics, ApJ, 738, 191

Brueckner, G. E., et al. 1995, Sol. Phys., 162, 357

Chen, P. F. & Shibata, K. 2000, ApJ, 545, 524

Chen, J. & Krall, J. 2003. JGR, 108, A11, SSH 2-1, CiteID 1410

Dennis, B. R., Zarro, D. M., 1993, The Neupert effect - What can it tell us about the impulsive and gradual phases of solar flares? Solar Phys., 146, 177

Emslie, A. G., Dennis, B. R., Shih, A. Y. et al., 2012, Global Energetics of Thirty-eight Large Solar Eruptive Events, ApJ, 759, 71

Feynman, J. & Martin, S. F. 1995, JGR, 100, 3355

Gilbert, H. R., Holzer, T. E., Burkepile, J. T., Hundhausen, A. J. 2000, Active and Eruptive Prominences and Their Relationship to Coronal Mass Ejections, ApJ, 537, 503

Gopalswamy, N. 1999, in Solar Physics with Radio Observations, Proceedings of the Nobeyama Symposium, held in Kiyosato, Japan, Oct. 27-30, 1998, Ed. T. Bastian, N. Gopalswamy and K. Shibasaki, (Nobeyama: Nobeyama Solar Radio Observatory), NRO Report No. 479, 141

Gopalswamy, N., 2010, Large-Scale Solar Eruptions, in Heliophysical Processes, eds. N. Gopalswamy, S. S. Hasan, and A. Ambastha, Springer, Berlin, p. 53

Gopalswamy, N., 2013, Observations of CMEs and models of the eruptive corona, in SOLAR WIND 13: Proceedings of the Thirteenth International Solar Wind Conference. Ed. G. Zank, AIP Conference Proceedings, AIP Publishing LLC, Melville, New York, Volume 1539, pp. 5

Gopalswamy, N., 2015, The dynamics of eruptive prominences, In: Solar Prominences, eds. J.-C. Viall and O. Engvold, Astrophysics and Space Science Library, Volume 415. Springer International Publishing Switzerland, 2015, p. 381

Gopalswamy, N., Lara, A., Kaiser, M. L., Bougeret, J.-L. 2001, JGR, 106, 2526112

**Figures**

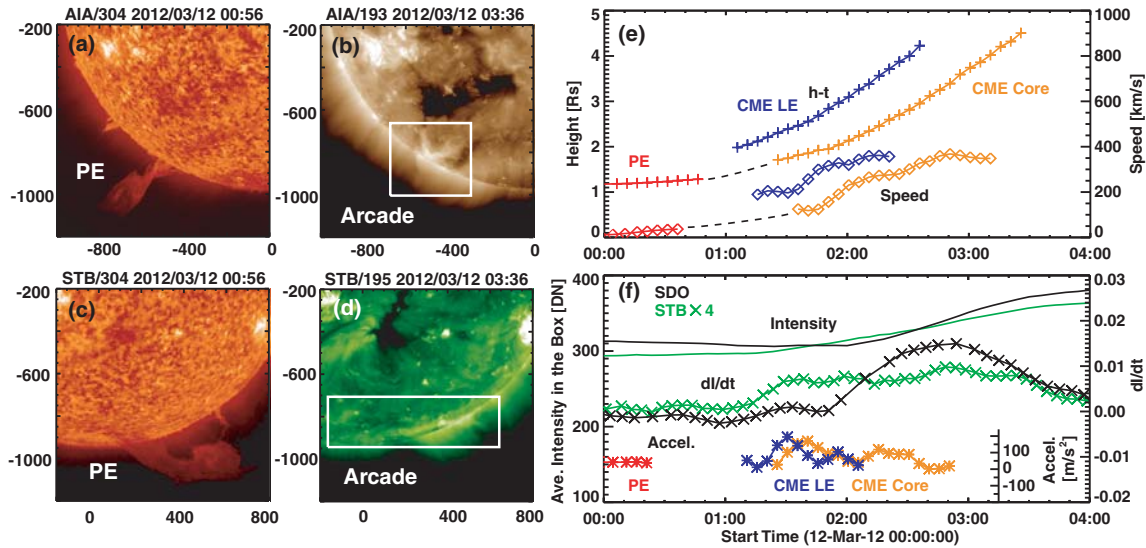

**Figure 1.** Prominence eruption (PE) from the polar crown (a,c) and the PEA (b,d) of the 2012 March 12 event as imaged by SDO and STEREO. PE is best observed in the 304 Å images, while the arcade is well observed in the 195 Å images. The heights and speeds of the CME and its core obtained from SECCHI/COR1 images are shown in (e) along with the corresponding measurements of the prominence (marked PE) from SDO/AIA images. The PEA intensity $I$ and its derivative $dI/dt$ are plotted from SDO/AIA 193 Å (black) and STEREO/EUVI 195 Å (green). The accelerations derived from the speeds in (e) for PE, CME, and CME core are also shown in (f). Note that the $I$ profile is similar to the CME speed profile in (e); compared with the acceleration profiles of the CME leading edge (LE) and the prominence core.

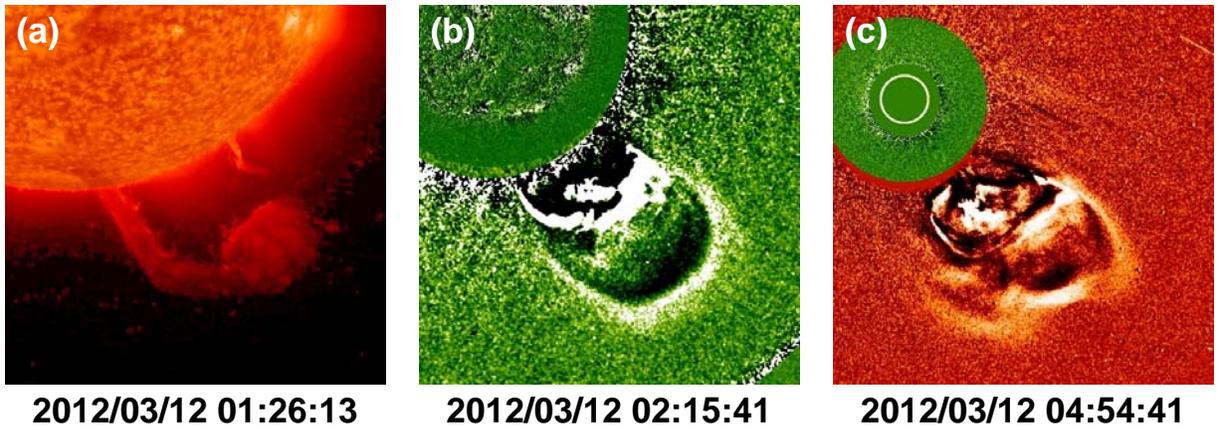
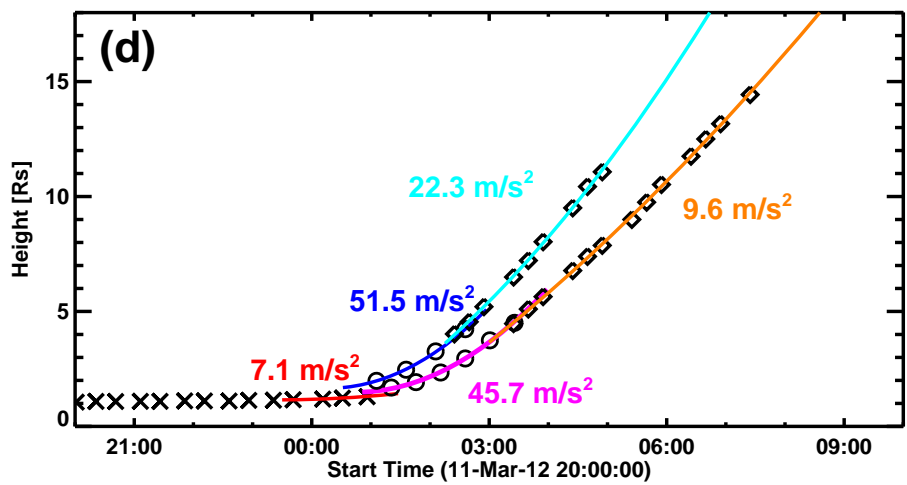

**Figure 2.** Snapshots of the eruptive prominence (a), CME in the STB/COR1 FOV (b), and CME in the STB/COR2 FOV(c). The composite height-time history of the prominence (lower curve) and the CME leading edge (upper curve) are shown in (d). In (d), the crosses refer to the prominence measurements from STB/EUVI; the circles and diamonds represent the measurements in the STB/COR1 and COR2 FOV, respectively. The average acceleration values within the three height ranges are indicated near the curves using second order fits.

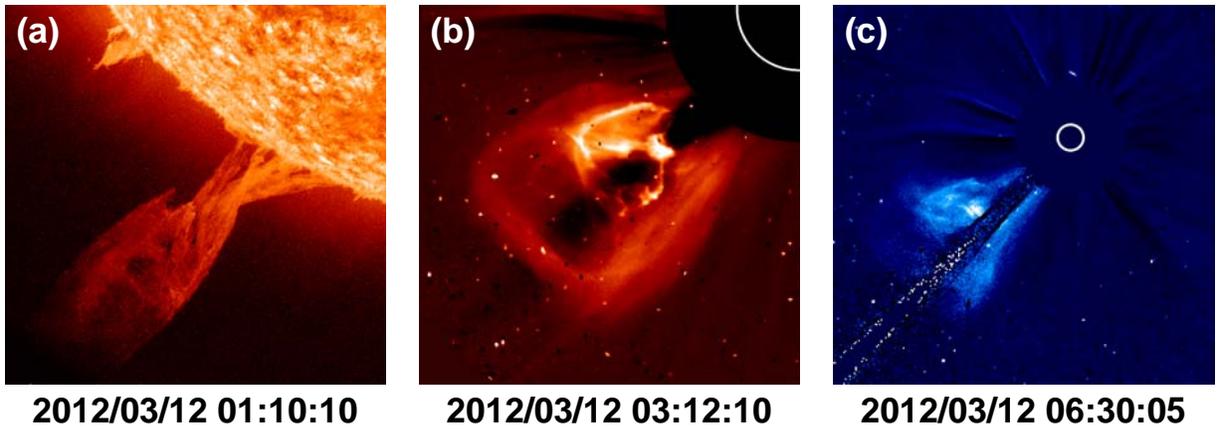
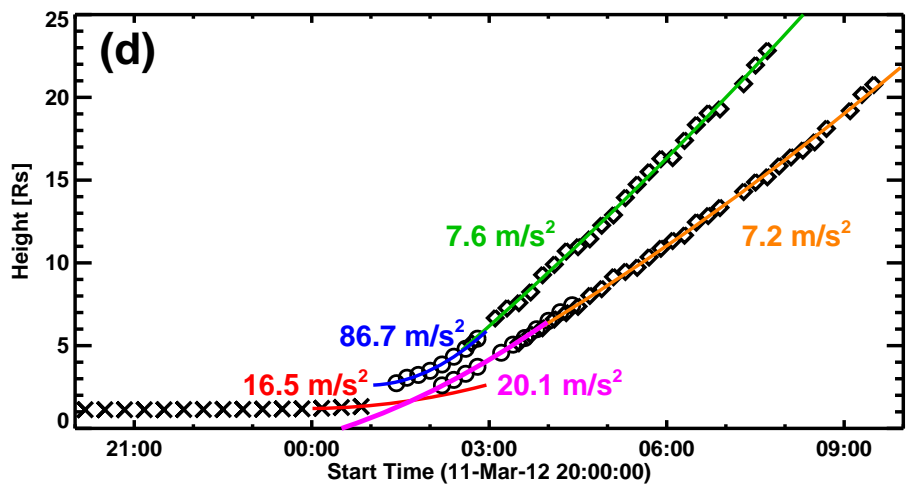

**Figure 3.** Snapshots of the eruptive prominence (a) observed by SDO/AIA I 304 Å; the CME in the LASCO/C2 FOV (b) and in the LASCO/C3 FOV (c). In the height-time plots in (d), crosses, circles, and diamonds represent SDO/AIA, LASCO/C2 and LASCO/C3 measurements, respectively of the CME leading edge (upper curve) and the prominence core (lower curve). The accelerations in the three height ranges are given near the curves.

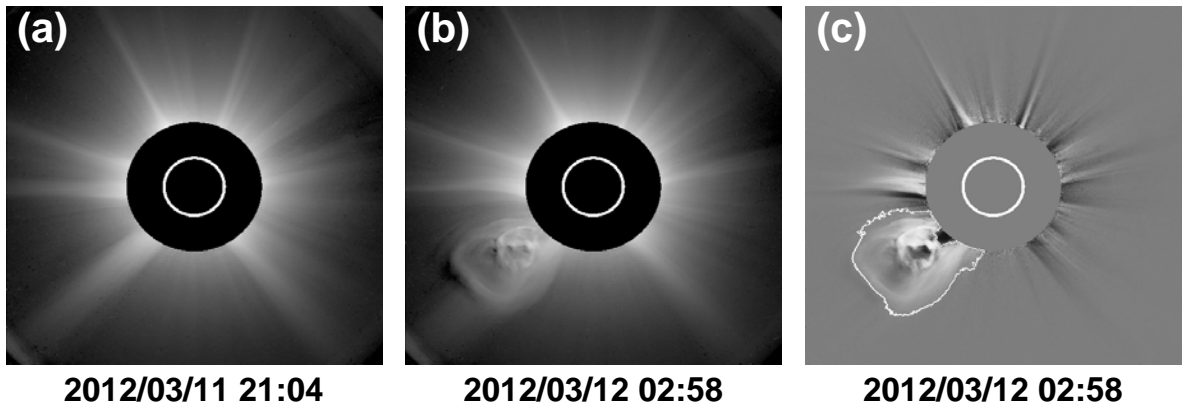

**Figure 4.** SOHO/LASCO polarized brightness (*pB*) images corresponding to the pre-event corona (a) and the CME (b). The difference image (c) was obtained by subtracting (a) from (b). The contour in (c) is at the level of $3.0 \times 10^{-11} B_o$, where $B_o$ is the mean solar brightness.

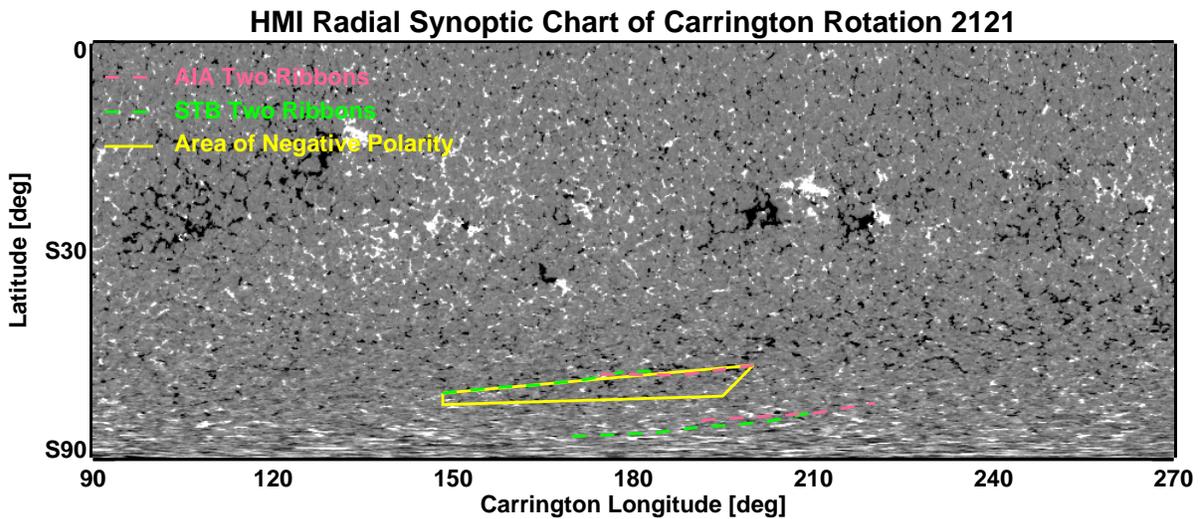

**Figure 5.** HMI radial synoptic chart for Carrington rotation 2121 (only the southern hemisphere is shown). White and dark patches represent positive and negative fields, respectively. The locations of the flare ribbons from SDO/AIA and STEREO/EUVI are represented by pink and green lines. The yellow quadrilateral represents the PEA on the negative polarity side. This is the region in which the magnetic field was estimated. We assume the source region to be the area covered by the PEA.

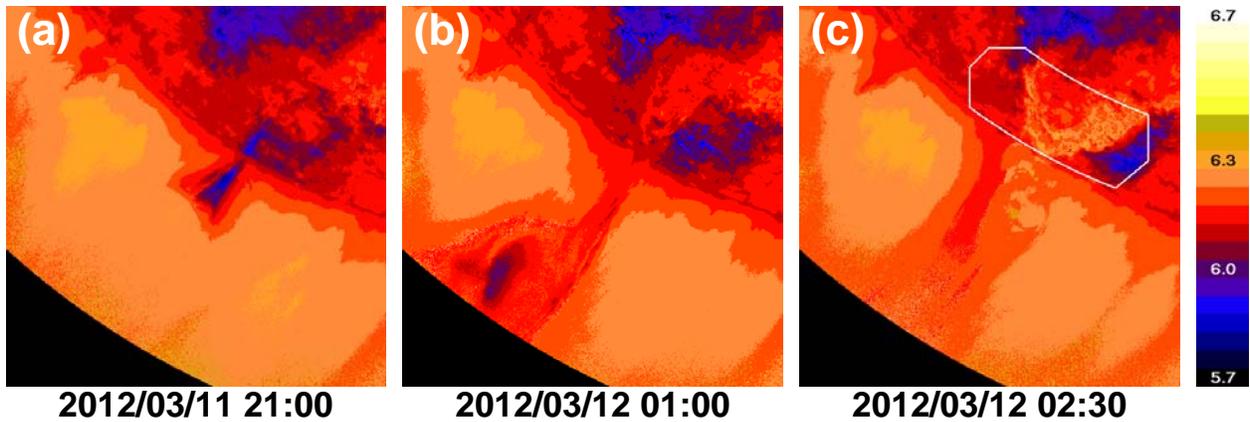

**Figure 6.** Temperature maps for the 2012 March 12 polar crown PE at three instances: (a) pre-eruption, (b) early phase of the eruption, and (c) late phase of eruption. The heated arcade can be seen in (c) within the polygon. Each temperature map was made from the 6 SDO/AIA EUV images at 94, 131, 171, 193, 211, 335 Å. The color bar to the right indicates the logarithmic temperature: 5.7, 6.0, 6.3, and 6.7 correspond to 0.5 MK, 1.0 MK, 2.0 MK, and 5.0 MK, respectively. Note that filament (the blue structure above the limb in (c)) and the coronal hole (at the top right in each panel) are below 1 MK, as expected.

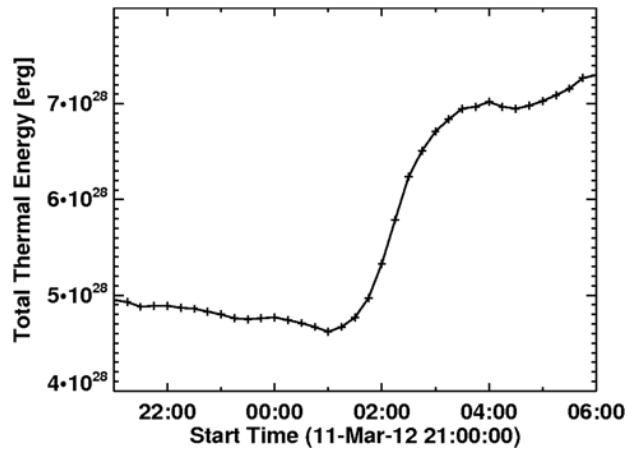

**Figure 7.** Peak thermal energy of the PEA as a function of time for the 2012 March 12 event. The highest value in the interval is $7.3 \times 10^{28}$ erg. This is about two orders of magnitude smaller than the peak thermal energy in X- and M-class flares associated with large CMEs.

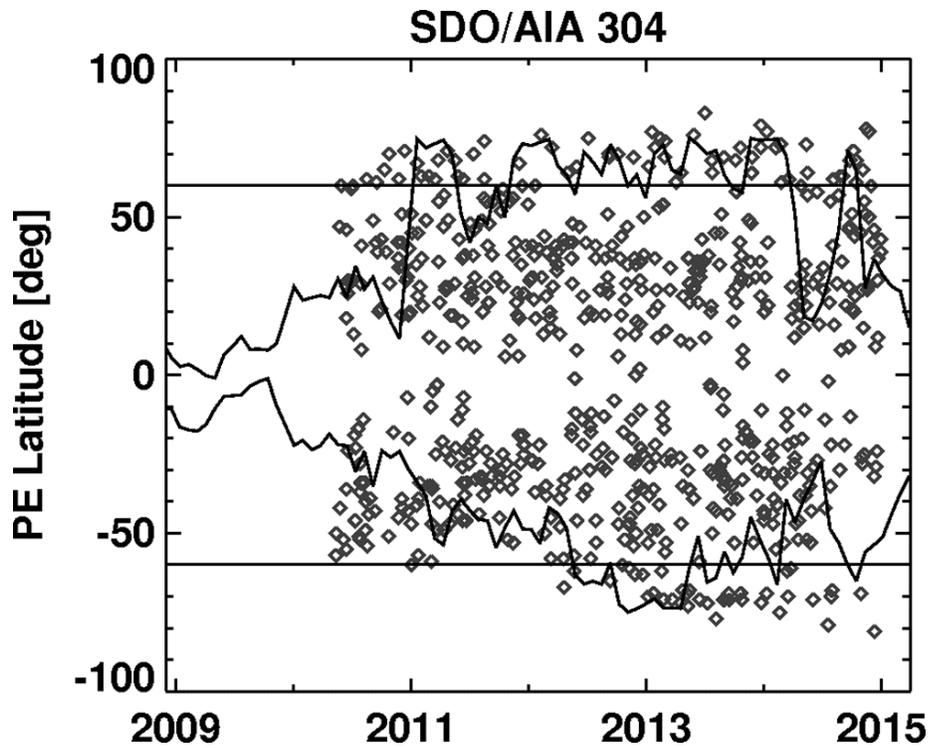

**Figure 8.** Time-latitude plot of PE sources obtained from SDO/AIA data. The PEs were detected automatically and then checked manually to eliminate noise events. The horizontal straight lines are at ±60º latitudes. The tilt angle of the heliospheric current sheet is overlaid in the northern and southern hemispheres (curves).

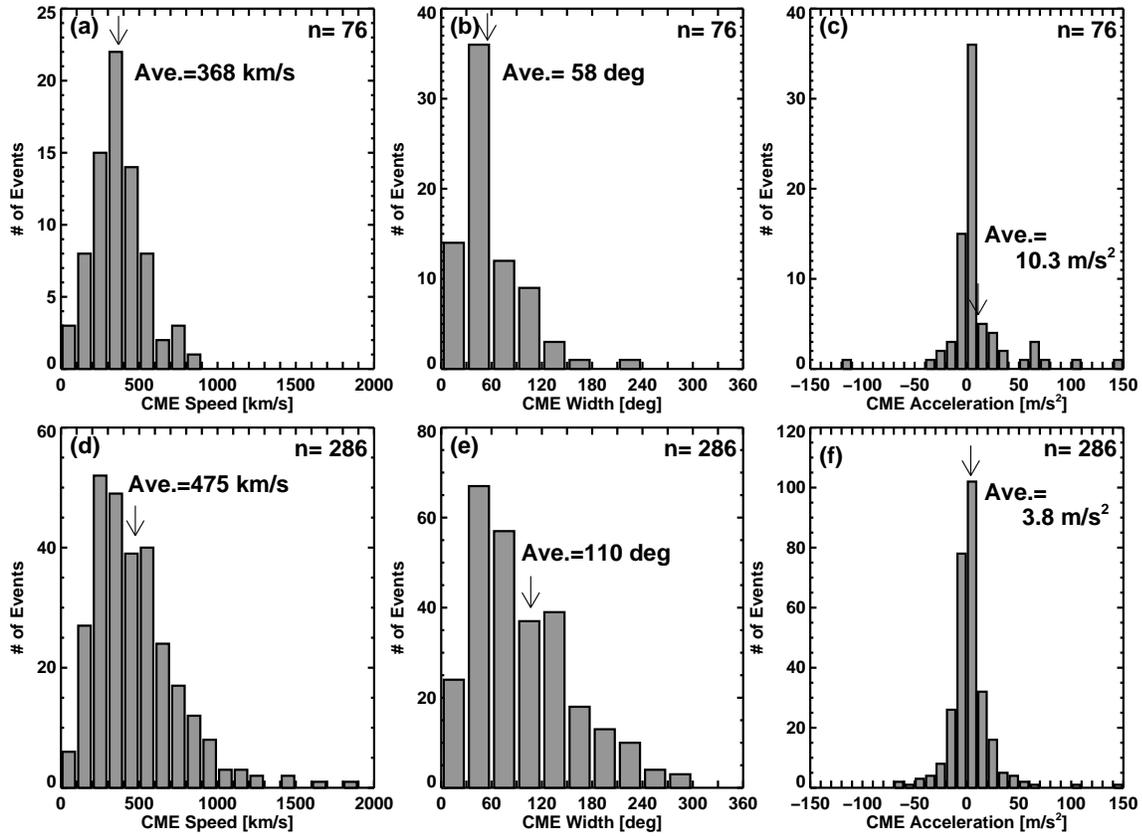

**Figure 9.** Speed, width, and accelerations of high-latitude (top) and low-latitude CMEs. For the high-latitude CMEs, we have restricted the latitudes to ≥60°; for low latitude CMEs, the latitudes are restricted to ±40°. Note that we did not separate active region and non-active region PEs at low latitudes. At high latitudes, all CMEs are associated with PEs from the polar crown.